\documentclass[twocolumn,pra,aps]{revtex4}
\usepackage{epsfig}

\begin{document}
\title{{\bf Vortex contribution to equilibrium currents}}
\author {Zeev~Vager}

\address{Department of Particle  Physics,\\
Weizmann   Institute of Science, 76100 Rehovot, Israel}

\date{ \today}

\begin{abstract}
For thin conducting rings, persistent currents were predicted to
depend periodically on the external magnetic field. It is shown
here that in general, whenever a periodic component of the current
is found, there must be a large aperiodic component forcing
aperiodic contribution to the equilibrium magnetic moment as a
function of the external field. It is also shown that contrary to
previous publications, the conduction electrons partition function
at low temperatures is always aperiodic as a function of the
external magnetic field. This finding is relevant to experimental
observations of magnetism of conducting rings as well as general
magnetization phenomena.

\end{abstract}
 \maketitle \bibliographystyle{unsrt}

\subsection{Introduction}
The general subject of persistent current in conducting rings is
geared around the notion of its being periodic with the magnetic
flux penetrating the rings. Many experimental measurements in
normally conducting rings\cite{ref1} confirm periodic components
with a fundamental flux period of $hc/e$ and its harmonics.
However, the experimental currents are significantly higher than
the theoretical predictions\cite{ref2}. In the theoretical
publications, the notion of periodicity is considered to be well
known, as if given by gauge invariance. The purpose of this letter
is to disagree with this notion and to show that a major aperiodic
component has been neglected. In superconducting magnets this is
evident since the persistent current is simply proportional to the
penetrating quantized flux. As will be shown, each time persistent
current exists in a ring, a large aperiodic component must persist
too.

The notion of magnetic flux periodicity of the partition function
of electrons within a superconductor ring, disregarding surface
currents, was coined by Byers and Yang\cite{BY}. Clearly, the
magnetic field of superconducting magnets does not show this
periodicity because it is solely due to surface currents. This is
not in contradiction with Byers and Yang which excluded electrons
in the penetration depth region of the superconductor. This
allowed Byers and Yang to use a restricted region electronic
Hamiltonian where the magnetic field is strictly set to zero.
Therefore, this restricted Hamiltonian cannot be used for
prediction of surface currents.

However this restricted Hamiltonian was used in the literature for
treating magnetization of rings. As a result of using the
restricted Hamiltonian, one obtains a misleading notion of
equilibrium current which depends periodically on the penetrating
magnetic flux. The present article indicates some of those
previous publications on equilibrium magnetization using such
incorrect Hamiltonian which resulted periodic behavior. The
leading article is by Bloch\cite{Bloch}, treating a
superconducting ring which contains a single barrier. Other
articles followed this periodic notion claiming similar results
for normal rings\cite{ref2,BIL}. As will be shown, an overlooked
electromagnetic energy in the Hamiltonian demands that whenever
equilibrium persistent current is measured in rings the dependence
on external magnetic field is always aperiodic. That does not
exclude a periodic component, as measured in numerous occasions.
The measurement of the periodic component of persistent current is
facilitated by observation of harmonics of the natural fundamental
component and therefore excludes the aperiodic component. This is
how the misleading notion of periodicity of persistent current got
its experimental fictitious support.

The correct approximation to the non-relativistic Hamiltonian and
its implications to magnetization is also highly relevant to the
understanding of experimental observations of magnetization that
were left unexplained so far. In particular, the mechanism of
giant magnetization of thin organic layers has been
disputed\cite{HER-GAR} by the use of the restricted Hamiltonian.
Therefore, revised considerations are needed.

\subsection{Preliminary remarks}

This article is about magnetization of conductors under special
conditions of equilibrium. At low enough temperatures, a
phenomenological difference between a normal conductor and a
superconductor is that the fraction of non-viscous conduction
electrons in a normal conductor tends to zero for infinite
length-dimensions while this fraction is independent of length in
a superconductor. It will be shown that the mere existence of
non-zero fraction of non-viscous electrons in a finite normal ring
is enough for equilibrium aperiodic magnetization.

In particular, consider a conductor ring in a uniform magnetic
field along the ring axis. Assume that there is a temperature
where a non-zero fraction of the conduction electrons flows
without viscosity (persistent current). This assumption leads to
aperiodic magnetization in both the classical limit and in quantum
mechanics.

The discussion starts with superconducting rings where flux is
quantized and persistent currents are simply proportional to the
trapped flux. In type I superconductors there are no transitions
between the quantized persistent currents for a finite range of
temperatures above zero. This is a case where the persistent
current is isolated even from the environmental radiation
temperature. In any finite ring, both type II superconductors and
normal conductors at low enough temperature have a non-viscous
fraction of conduction electrons, sometimes, ever so small but
non-zero. Ignoring radiation, the superconducting ring
quantization of the non-viscous fraction is valid. Including
radiation, equilibrium is established and the quantized currents
have statistical weights resulting a major current component which
is linear with the external field. The remaining periodic current
component is generally smaller. Detailed discussion is given
below.

\subsection{Superconducting rings}
In 1961, a celebrated experiment by Deaver and Fairbank\cite{Fair}
had shown that superconducting hollow cylinders sustain quantized
magnetic flux $\phi_n$ which are integer, $n$, multiples of
$hc/2e$. From that time it is accepted that superconducting rings
carry currents which are proportional to the trapped quantized
flux. If such rings are held in place (in relation to
environment), the far magnetic-dipole field is quantized too. The
quantization is known to hold within some limited range of
temperatures and external magnetic fields. The flux dependent
energy of a superconducting ring within a uniform external
magnetic field $B$ is given by
\begin{equation}\label{FE}
    E_n(B)=\frac{1}{2{\cal L}}\phi_n^2-\frac{1}{{\cal L}}AB\phi_n
    \equiv
    \frac{1}{2{\cal L}}\phi_n^2-\frac{1}{{\cal L}}\Phi\phi_n
\end{equation}
where $\cal L$ is the inductance of the superconducting ring and
$A$ is an effective area\cite{footn1} related to the ring. When
the magnetic dipole of a superconducting ring is pointing along
the external magnetic field, $A$ is maximal and $\Phi$ is the
external flux through that area without the superconducting ring.

Plugging in the flux quantization, the energy expression is
\begin{equation}\label{FE2}
    E_n(B)=\frac{\hbar^2}{2\cal I}(n-\eta)^2
    -\frac{\hbar^2}{2\cal I}\eta^2
\end{equation}
where
\begin{equation}\label{I}
    {\cal I}=4\pi^2 \alpha^2 m_0a_0{\cal L}
    =4\pi^2 m_0r_0{\cal L}\quad\mbox{and}\quad
    \eta=\frac{\Phi}{hc/2e}
\end{equation}
($\alpha$ - fine structure constant, $m_0$ - electron mass, $a_0$
- Bohr radius and $r_0=e^2/m_0c^2$). These are shifted rotational
states with large rotational energies. Since the moment of inertia
of a single electron on the cylinder is ${\cal I}_0=m_0({\cal
L}/2\pi)^2$ then
\begin{equation}\label{Ies}
    {\cal I}={\cal I}_0/K
\end{equation}
where
\begin{equation}\label{K}
    K=\frac{1}{(4\pi)^4}\frac{\cal L}{r_0}\,.
\end{equation}
Even for atomic scale and especially for macroscopic or mesoscopic
rings, $K$ is a huge number. Conversely, the effective moment of
inertia is by far smaller than ${\cal I}_0$. Most importantly, the
phenomenon is independent of the non-viscous fraction of
conduction electrons as long as it is non-zero.

The $e^{-2}$ energy cofactor in (\ref{FE2}) forces the need of
non-perturbative solutions of the fields. Therefore, use of
electromagnetic potentials within hamiltonian formulations should
be treated with caution\cite{Dirac}. This is elaborated next.
Lastly, it should be clear that the last term
$-\hbar^2\eta^2/2{\cal I}=-A^2B^2/2\cal L$ is of classical origin
and signify the reduction of the overall magnetic energy for any
quantum state.

\subsection{Formal note on non-relativistic hamiltonian}

A non-relativistic approximation to the classical hamiltonian for
charged particles in an electromagnetic field is
\begin{equation}\label{CH}
    {\cal H}=\sum_n
    \frac{1}{2m_n}(\vec{p}_n-\frac{e_n}{c}\vec{A})^2+U
    +\frac{1}{8\pi}\int{(B^2+E^2)}d\tau
\end{equation}
where $(U,\vec{A})$ is the relativistic vector potential which
defines the fields $\vec{B},\vec{E}$. It is understood that this
vector is defined up to a gauge transformation. Such
transformations do not change observations or fields. The total
electromagnetic energy is a function of charge and current
density. Therefore, the last term turns to an operator in the
quantum Hamiltonian.

Certainly, the eigenvalues in (\ref{FE2}) are eigenvalues of such
a hamiltonian. Adding to the last term in (\ref{FE2}), the
positive and much larger constant $\int{B_0^2}d\tau$ (here $B_0$
is the external field), completes the integration on the energy of
the electromagnetic field. As mentioned above, the eigenvalues
(\ref{FE2}) result from nonperturbative treatment of the equations
of motion. Specifically, the electromagnetic field cannot be
treated as a perturbation.

The last term in (\ref{FE2}) cannot be derived from a hamiltonian
of the form
\begin{equation}\label{CH1}
    {\cal H}=\sum_n
    \frac{1}{2m_n}(\vec{p}_n-\frac{e_n}{c}\vec{A})^2+U
\end{equation}
since the classical reduction of the magnetic energy (see above)
is missing.

Whenever equilibrium current exists in a loop, there is a magnetic
dipole far field and the associated flux lines can be traced going
in and out the current boundary. The current and this flux are
proportional to each other with a constant of proportionality
which depends on the geometry of the loop. This is a classical
relation which stems from Maxwell equations. The function of the
dipole self energy together with its interaction with external
field has a minimum at a non-zero dipole moment.

As will be explained below, topologically, this situation is
consistent with an Abrikosov-Nielsen-Olesen vortex. The far dipole
field defines an axis along which north and south poles reside. As
shown by Wu and Yang\cite{WY}, for any vector potential
reproducing the dipolar magnetic field there is, at least, a
section on the dipolar axis where the vector potential is
undefined. Different vector potentials which reproduce the dipolar
magnetic field and are well defined on that section are undefined
on, at least, another section along the axis. Since the current
carrying charged particles do not pass through the axis connecting
the poles, any of the vector potentials reproducing the
electromagnetic fields can be used within the hamiltonian provided
that it explicitly contains the electromagnetic energy operator.
It is concluded that the quantum hamiltonian of circulating
current in a loop must include the magnetic energy like
(\ref{CH}).

\subsection{Aharonov-Bohm flux and conducting rings}

There is a well known theorem that if a magnetic flux $\phi$
penetrates through a conducting ring without passing through the
conductor (Aharonov-Bohm (AB) flux), all equilibrium physical
properties of the ring are periodic with $\phi$. The period is
given by $hc/e$. Under such conditions, if there is an equilibrium
current then it is periodic with $\phi$.

How does one understand the currents in superconducting rings
where persistent currents are aperiodic by simply being
proportional to the flux within the ring? The answer is that there
is a thin inside layer where the magnetic field penetrates and
interacts with the circulating current. This is topologically
equivalent to a flux line penetrating a type II superconductor
accompanied by a super-current starting at zero at the flux line
and decaying exponentially after a small radial distance (compared
with the macroscopic dimensions of the superconductor). Such
phenomenon is known as Abrikosov-Nielsen-Olesen vortex line which
is a soliton solution to the relativistic field equation. In a
superconducting ring, the current in the thin penetration depth
and the flux within the ring interact with each other to form a
'fat' Abrikosov-Nielsen-Olesen vortex 'line'. It is \textbf{NOT}
an AB flux which does not penetrate the conductor by definition.

Conversely, if an 'infinite' thin coil carrying a flux $\phi_{AB}$
is inserted to any loop, its positive self energy is proportional
to $\phi_{AB}^2$. If the loop is superconducting and no other flux
penetrates it then there is an additional quantum term to the
total energy which is $\hbar^2(n-\eta_{AB})^2/2\cal I$. There is
no reduction of the magnetic energy due to interaction of the
dipole field in the external field.

In a ring conductor having non-viscous flow, not necessarily
superconducting, flux penetrates the conductor as well as the hole
in the ring. Eigenenergies like (\ref{FE2}) are a sum of two
terms. The first term is proportional to $(n-\eta)^2$ and behaves
as if it is an AB flux. The second term is of classical origin and
much larger, except near zero flux. This is further explained
below.

\subsection{Equilibrium} When leakage of flux is possible (such
as in type II superconductors), equilibrium can be achieved with
the thermal radiation. The large rotational energy factor in
(\ref{FE2}) selects Boltzmann statistical factors such that the
equilibrium is very near the one quantized flux which is nearest
to $\Phi$ (where $|n-\eta|$ is minimal). Thus, at equilibrium,
paramagnetic magnetization is reached. The partition function is
\textbf{NOT} periodic with $\eta$. As will be discussed later,
this aperiodic dependence on $\Phi$ contributes a smooth behavior
of the partition function. If the parabolic classical energy
function\cite{footn2}
\begin{equation}\label{cf}
    \bar{E}(\eta)=-\frac{\hbar^2\eta^2}{2\cal I}=
    -\frac{A^2}{2\cal L}B^2
\end{equation}
is artificially subtracted from (\ref{FE2}) then it becomes
periodic with $\eta$. But this energy part of (\ref{FE2}) is
clearly paramagnetic. Its simple physical interpretation is that
non-zero equilibrium persistent current creates a magnetic dipole
which minimizes the self energy of the dipole together with the
dipole interaction energy. It is interesting to compare it with
deductions of Bloch\cite{Bloch} were the parabolic dependence is
missing. As mentioned above, this is just one of many references
where persistent current is considered as periodic with the
external field.

\subsection{Normal conductor rings}
The extension to normal conductors is realized by recognizing that
once a non-zero fraction of the conduction electrons is
non-viscous, the soliton energy dependence remains exactly as
above. The only difference from type II superconductors is the
rate of relaxation to equilibrium. The classical arguments are as
follows.

Consider a normal ring in an external uniform magnetic field $B$
which is in the direction of the ring axis. If a current $I$
circulates in the ring then the current dependent energy of the
system is
\begin{equation}\label{EI}
    U=\frac{1}{2}{\cal L}I^2-ABI
\end{equation}
which is the classical analogue of (\ref{FE}). The main difference
here is that the effective area $A$ is significantly larger than
the inside hole, yet smaller than the total area. An immediate
consequence is that at thermal equilibrium (with radiation) there
is a mean current
\begin{equation}\label{I0}
    I_0=AB/{\cal L}\,.
\end{equation}
In terms of $I_0$ the energy dependence on $I$ is
\begin{equation}\label{EI1}
    U=\frac{1}{2}{\cal L}(I-I_0)^2-\frac{1}{2}{\cal L}I_0^2
\end{equation}
which is negative at the vicinity of $I_0$. It is concluded that
the ring is paramagnetic in nature. Moreover, the magnetic moment
at thermal equilibrium is given by
\begin{equation}\label{mm}
    m=-\frac{\partial U}{\partial B}|_{I_0}
    =\frac{A^2}{\cal L}B\,.
\end{equation}
This is the classical magnetic moment of the ring in an external
field $B$, provided that there is a nonzero fraction of
non-viscous conducting electrons. This could be measured by a
magnetometer. The equilibrium energy of the ring is lowered by the
existence of a magnetic field $B$ by the negative energy
\begin{equation}\label{mB}
    -m B=-\frac{A^2}{2\cal L}B^2=-\frac{\Phi^2}{2\cal L}\,.
\end{equation}
This energy is just the classical function in (\ref{cf}).

It should be noted that whatever quantum considerations dictate,
the correspondence principle require that in the classical limit,
the free energy dependence on $I$ must be of form (\ref{EI1}).
This is indeed consistent with the behavior of superconducting
rings, as in (\ref{FE2}).

\subsection{Conclusion}

If the interest is restricted to circulating currents in rings, an
effective hamiltonian can be formulated which includes the soliton
energy and the correct dependence on external magnetic field. The
eigen-energies for a superconducting ring are given by
(\ref{FE2}). When the Meissner effect is not applicable but
non-viscous currents exist, the only perturbation comes from
thermal radiation. Therefore, Boltzmann factors can be define for
each eigen-energy and the partition function is well defined.

Each eigen-energy is a sum of two terms\newline
$E_n(B)=E^{(Q)}_n(B)+E^{(C)}(B)$ where
\begin{eqnarray}\label{EQ}
    E^{(Q)}_n(B)=\frac{\hbar^2}{2\cal
    I}(n-\eta)^2\quad\mbox{and}\\
    \quad E^{(C)}(B)=-\frac{A^2}{2\cal L}B^2\,.
\end{eqnarray}
Thus, the partition function, $Z$, has a common factor:
$\exp[-\beta E^{(C)}(B)]$. Two terms contribute to the
magnetization when the derivative of $Z$ is taken with respect to
$B$: a stable classical term (\ref{mm}) and a term where a
statistically weighted sum of alternating sign
magnetization\cite{footn3}. The second term contribution\cite{BIL}
depends on details of the conductor and its theoretical
predictions underestimate\cite{ref2,vag} experimental
results\cite{ref1} of the periodic component. The first classical
term is quite illusive and, so far, found only indirectly on
molecular monolayers\cite{RZ}. Obviously, such magnetization
appear in any current loop where losses are compensated by power
supplies.

The 'classical' magnetic energy which must be added to the
eigenenergies of the hamiltonian (\ref{CH1}) is the additional
'classical' energy of the separable hamiltonian (\ref{CH}).
Similarly, the current amplitude is a superposition of current
amplitudes of those two separable parts.

This article emphasizes the importance of the previously ignored
classical magnetic energy to the free energy of conduction
electrons. The phenomenon of aperiodic magnetization should appear
in any of the persistent current experiments and many other
magnetization phenomena.

Illuminating discussions with Prof. A. Schwimmer as well as a
lesson in classical field theory are acknowledged.


\begin{thebibliography}{99}
\bibitem{footn1} An effective area can be chosen so that the energy is
given by (\ref{FE}). The electromagnetic units are chosen such
that $\cal L$ has length units.
\bibitem{footn2} For the total energy, this negative energy is added to
the much larger
positive homogeneous magnetic energy of the external field in
order to get the full energy of the system. The condition
$A^2/{\cal L}\ll V$ is compatible with the 'homogeneous field'
demand, where $V$ is the volume of the field.
\bibitem{footn3} The order of the energy levels is given by the
ordered $(n-\eta)^2$ therefore the derivative with respect to $B$
has alternating signs for ordered energy levels.
\end{thebibliography}
\end{document}